\documentclass[prb,preprint,aps,showpacs]{revtex4}
\usepackage{graphicx}

\input epsf.sty
\newcommand{\norm}[1]{\left\Vert#1\right\Vert}
\def\({\left(}
\def\){\right)}

\begin{document}

\title{Statistics of transmission in one-dimensional disordered systems: universal characteristics of states in the fluctuation tails.}

\author{L. I. Deych}
\author{M. V. Erementchouk}
\author{A. A. Lisyansky}
\affiliation{Queens College of City University
of New York, Flushing, NY 11367}
\author{Alexey Yamilov}
\author{Hui Cao}
\affiliation{Physics Department, Northwestern University}
\begin{abstract}
We numerically study the distribution function of the conductance
(transmission) in the one-dimensional tight-binding Anderson and
periodic-on-average superlattice models in the region of
fluctuation states where single parameter scaling is not valid. We
show that the scaling properties of the distribution function
depend upon the relation between the system's length $L$ and the
length $l_s$ determined by the integral density of states. For
long enough systems, $L \gg l_s$, the distribution can still be
described within a new scaling approach based upon the ratio of
the localization length $l_{loc}$ and $l_s$. In an intermediate
interval of the system's length $L$, $l_{loc}\ll L\ll l_s$, the
variance of the Lyapunov exponent does not follow the predictions
of the central limit theorem and this scaling becomes invalid.

\end{abstract}

 \maketitle

\section{Introduction} Coherent transport properties of
disordered systems have been a subject of active research for the
last thirty years, but complete understanding of this phenomenon
even for one-dimensional models is still absent. Even though the
scaling theory, put forward in the pioneering work of
Ref.~\onlinecite{gang}, created a successful conceptual framework
for discussing the phenomenon of localization, the theoretical
foundation of the scaling hypothesis itself has not yet been
completely understood.  One of the principal difficulties that the
scaling theory of localization had to deal with from the very
beginning was an absence of self-averaging of the main transport
coefficients: conductance, $g$, or transmission, $T$. Therefore,
even the nature of the scaling parameter remained unclear until it
was realized that the scaling hypothesis has to be applied to the
entire distribution function of the conductance or
transmittance.\cite{Anderson,Abrikosov,Shapiro,LGP}

For one-dimensional systems Anderson, et al.\cite{Anderson}
suggested that the most suitable quantity for dealing with the
statistical description of conductance is the Lyapunov exponent
(LE), which can be defined for systems with finite length $L$ as
\begin{equation}\label{LE}
\tilde{\gamma}=-\frac{1}{L}\log\left(1+\frac{1}{g}\right)
=-\frac{1}{L}\log T
\end{equation}
The name `Lyapunov exponent' alludes to the fact that the quantity
defined by Eq.~(\ref{LE}) have the same statistical properties as
the ``real'' Lyapunov exponent, i.e. the exponential growth rate,
$(1/L)\log|\psi|$, of the norm of the wave function, $\psi$. An
important property of LE is that it satisfies a multiplicative
central limit theorem\cite{Oseledets} and approaches a non-random
limit, $\gamma$, when the size of the system, $L$, tends to
infinity. The localization length, $l_{loc}$, of a state with
energy $E$ in the infinite system is related to $\gamma$ as
$l_{loc}=\gamma^{-1}$. At finite $L$, $\tilde{\gamma}$ is a random
quantity with mean value equal to $\gamma$,
$\langle\tilde\gamma\rangle = \gamma$. The distribution of LE is
the main object of research in the field of one-dimensional
localization. The hypothesis of single parameter scaling (SPS) in
this context means that the distribution function can be
parameterized by a single parameter, $\gamma$ itself. As a result,
it is expected that all moments of the distribution can be
expressed in terms of the first moment,
$\langle\tilde\gamma\rangle$, in a universal way. For the second
moment (variance), $\sigma^2$, such a relationship, as it was
first conjectured by Anderson et al.,\cite{Anderson} can be
presented in the form
\begin{equation}\label{SPS}
    \sigma^2=\frac{\gamma}{L},
\end{equation}

The entire distribution function of LE for systems with finite
lengths was also derived by several authors in the limit of
infinitesimally weak local scattering for several
models.\cite{Abrikosov,DPMK,Mello} For finite $L$, this function
was found to be non-Gaussian, but nevertheless, it depended upon a
single parameter - the localization length.

Thus, in the situations when SPS holds the problem of the
conductance/transmision distribution function can be considered as
settled. There are spectral regions, however, where SPS fails even
for locally weak disorder. These are, first of all, the regions of
fluctuation states, which arise outside of the initial spectrum
because of disorder. This result was first obtained numerically in
Ref.~\onlinecite{Zaslavsky} for a periodic-on-average system and
was confirmed by an exact analytical solution of the Lloyd model
(the Anderson model with the Cauchy distribution of the site
energies).\cite{DeychPRL2000,DeychPRB} Similar results were
obtained numerically for the Anderson model with the
box\cite{DeychPRL2000,DeychPRB} and dichotomic\cite{Erementchouk}
distributions of the site energies, and analytically for a
continuous model with white noise Gaussian
potential.\cite{Schomerus} The analytical calculations of
Refs.~\onlinecite{DeychPRL2000,DeychPRB} revealed that the
criterion for the validity of SPS can be presented in the form
$l_{loc}>l_s$, where $l_s$ is a new scale introduced in
Refs.~\onlinecite{DeychPRL2000,DeychPRB}. For the Lloyd model this
scale is defined in terms of the imaginary part of the Lyapunov
exponent, which, according to Thouless\cite{Thouless} is
proportional to the integral density of states. Therefore, $l_s$
can be presented in the form
\begin{equation}\label{ls}
    l_s=\frac{1}{\sin{\left(\pi N(E)\right)}},
\end{equation}
where $N(E)$ is the integral density of states between the genuine
boundary of the spectrum and the energy $E$ normalized by the
total number of states in the system. The definition of $l_s$ in
this form can be easily generalized to other models as well, and
it was shown numerically that the SPS criterion based upon $l_s$
works for such models as the Anderson model with
box\cite{DeychPRL2000,DeychPRB} and dichotomic\cite{Erementchouk}
distributions of site energies, a Kronig-Penney-like model with a
periodic-on-average distribution of barrier
widths,\cite{DeychPRL2000,DeychPRB} and a model of a scalar wave
propagating in an one-dimensional absorbing disordered
medium.\cite{Yamilov} The case of periodic-on-average models
involves a system with multiple bands, and in this case $N(E)$
must be understood as the integral density of states between a
genuine boundary of the band (if the latter exist) and the energy
$E$ normalized by the total number of states in the band. More
detailed discussion of this case can be found in
Ref.~\onlinecite{DeychPRB}. In a recent paper
Ref.~\onlinecite{COBanomaly}, it was shown how this criterion can
be applied to the zero energy states of the Anderson model with a
diagonal disorder, where the violation of SPS was observed in
Ref.~\onlinecite{Titov}.

The criterion based on $l_s$ replaces an original criterion put
forward by Anderson, et al.\cite{Anderson} that suggested that SPS
exists if the stationary distribution of the phases of the
reflection and transmission coefficients is uniform, and the phase
relaxes to this distribution over a length, which is much smaller
than the localization length. By using the hypothesis of the phase
randomization Eq.~\ref{SPS} was re-derived by many authors for a
variety of different models.\cite{rederivations} The phase
randomization was proven rigorously in some
one-dimensional\cite{Derrida,Goldhirsch,LGP} and
quasi-one-dimensional\cite{Mirlin1,Mirlin2} models, but only for
certain parts of the spectrum of the respective systems. At the
same time, it was found that, for instance, in the Anderson model
with a diagonal disorder the stationary distribution of the phase
is not uniform for all values of energy $E$, for which
$\cos^{-1}{(E/2)}$ is a rational fraction of $\pi$ (it is assumed
that in the non-random case all site energies in the Anderson
model are set to zero, and the interaction parameter is chosen to
be equal to unity). The strongest deviation of the phase
distribution from the uniform one takes place in the vicinities of
$E=0$ and the initial band boundaries $E=\pm 2$.  While it was
found that an absence of the phase randomization in both of these
cases is accompanied by the violation of
SPS,\cite{Zaslavsky,DeychPRL2000,DeychPRB,Titov} the reference to
the phase randomization as a criterion for SPS does not seem to be
satisfactory. Indeed, the initial idea of the phase randomization
length,\cite{Anderson} used to introduce the criterion for SPS,
does not actually describe the way the distribution of phase
becomes non-uniform. The absence of the phase randomization does
not  mean that the relaxation length of the distribution of phase
becomes too large and exceeds the localization length. What it
means is that  the stationary distribution of phase, which can be
reached over relatively short distances, is merely not uniform.
Thus, the problem of a criterion for SPS is simply replaced by the
problem of finding a  criterion describing the transition between
uniform and non-uniform stationary distributions of the phase. A
solution for the latter problem suggested, for instance in
Ref.~\onlinecite{Titov}, applies to only one particular model,
and, actually involves
 different criteria for different spectral regions.
In contrast, the criterion based on $l_s$ introduced in
Refs.~\onlinecite{DeychPRL2000,DeychPRB} was proven to work for
the entire spectrum of the variety of different models, and
offers, therefore, a  universal approach to the verification of
SPS.

The violation of SPS in the spectral region of fluctuation states
rises a question about the properties of the probability
distribution of LE in these regions. Recently, a significant
progress in this direction was achieved in
Refs~.\onlinecite{Schomerus,DeychPRL2003}. In the former paper,
the first four moments of this distribution were found
analytically for the Anderson model with a Gaussian white-noise
potential.  The authors of the latter paper used numerical
simulations to develop a macroscopic scaling approach to this
problem, and one which could be readily applied to a wide variety
of different systems.  It was shown in
Ref.~{\onlinecite{DeychPRL2003}, that not only second, but also
the third moment of the distribution function of LE for the
Anderson tight-binding model with diagonal disorder can be fully
characterized by a scaling parameter $\kappa=l_{loc}/l_s$.

The objective of the present paper is to present more fully and
expand the results of Ref.~{\onlinecite{DeychPRL2003}. Considering
two quite different models of one-dimensional localization such as
the Anderson tight-binding model with a diagonal disorder, and a
model of a scalar wave propagating in a one-dimensional random
superlattice, we  demonstrate that one-dimensional disordered
systems allows for a universal scaling description of the
conductance (or transmission) distribution in the spectral regions
of fluctuation states, where standard SPS does not work. In
particular, we show that the scaling approach suggested in
Ref.~{\onlinecite{DeychPRL2003} describes not only the Gaussian
bulk of the distribution function, but is also capable of
describing the statistics of large deviations characterized by the
third moment of the distribution.

The results presented in this paper are also relevant to the
problem of resonant tunnelling through disordered potential
barriers. For the first time, this problem was considered in the
pioneering work by Lifshits and Kirpichenkov\cite{Lifshits} for
quantum particles incident on a three-dimensional barrier, and was
later studied in many subsequent papers (see reviews in
Ref.~\onlinecite{Pollak,Raikh}). Mostly, these works were
concerned with tunnelling through 3-dimensional barriers with the
dimension in the  propagating direction much smaller than in the
perpendicular directions. Even though the resonant tunnelling is
in many aspects a quasi-one-dimensional process,\cite{Lifshits}
the transport in the pure one-dimensional models significantly
differs from the situation described above. First of all, in
one-dimensional case all states are localized and transmission at
any energy can be described as a resonant under-barrier
tunnelling. Therefore, the difference between transport in the
region of states from the initial allowed bands and the
fluctuation states, is not as clear as in three-dimensional
situations. Therefore, the problem of transport via fluctuation
states was not considered as a separate problem in the area of
one-dimensional localization until very
recently.\cite{DeychPRL2000,DeychPRB} Second, the main quantity of
interest in the case of three dimensional barriers is the total
transmittance across the entire area of the barrier, which is
determined by the sum of individual transmissions through
independent quasi-one-dimensional channels or
filaments.\cite{Lifshits,Raikh} This quantity approaches a
non-random limit when the area of the barrier tends to infinity.
In a pure one-dimensional case the self-averaging quantity is the
Lyapunov exponent, which becomes non-random when the length of the
system becomes infinite. In a sense, the pure one-dimensional case
is an opposite limit to the one considered for three-dimensional
barriers. At the same time, solutions of the one-dimensional
problem can be used to describe barriers whose lengths are larger
than the typical localization length of individual channels.

Another important application of the problem studied in this paper
lies in the field of random lasing, which has become an area of
active research.\cite{random lasing} It is anticipated that using
localized modes of a strongly scattering disordered medium, one
can obtain very low-threshold lasing.  Disordered photonic
crystals, which support fluctuation photon states in the band-gaps
of the underlying periodic structures, can play an important role
in achieving this objective.\cite{vardeny,disphc} The results
presented in this paper will help to understand the unusual
statistical properties of the lasing threshold and the nature of
lasing modes in such structures.

\section{Models and technical details}

In this paper we study two models of one-dimensional Anderson
localization: a classical Anderson tight-binding model with a
diagonal disorder, and a scalar wave propagating in a
one-dimensional random superlattice. The Anderson model is
described by the equation of motion
\begin{equation}  \label{eq:An eq_of_motion}
\psi_{m+1} + \psi_{m-1} + (U_m - E)\psi_m = 0,
\end{equation}
where random on-site energies $U_m$ are described by a uniform
probability distribution:
$$
P(U_m) = \left\{\begin{array}{cc}\displaystyle{\frac{1}{2U}}, &|U_m| < U\\
                                   0,&|U_m| > U
                                   \end{array}\right.
$$
The propagation of a scalar wave is described by a regular wave
equation,
\begin{equation}\label{eq:Wave_eq_of_motion}
    \frac{d^2\psi}{dx^2}+k^2\epsilon(x)\psi=0,
\end{equation}
with a piece-wise dielectric  function, corresponding to a
superlattice consisting of two types of layers with dielectric
constants $\epsilon_1$ and $\epsilon_2$, respectively. The width
of the layers of the first kind is kept constant and is equal to
$d_1$, while the width of the layers of the second type was chosen
from a random distribution. In this paper, we report the results
for ({\it i}) $d_2$ uniformly distributed in the interval $\langle
d_2\rangle -\delta,\langle d_2\rangle +\delta$ (uniform
distribution), ({\it ii}) $d_2$ taking one of two equally probable
values $-\delta/\sqrt{3}$ and $+\delta/\sqrt{3}$ (dichotomic
distribution).

Both these models can be studied using the transfer matrix
approach, in which the propagation of the excitation along the
system is presented in the following form
\begin{equation}\label{eq:TMgeneral}
v_{m+1}=T_mv_m,
\end{equation}
where $v_m$ is a two dimensional state vector, which presents the
state of the system at the $m$-th site (or $m$-th interface
between the layers) and $T_m$ is the transfer matrix describing
the change of this state at one discreet step.  For the Anderson
model the state vector and the transfer matrix have the following
forms respectively
\begin{equation}\label{AMv}
v_m=\left(\begin{array}{c}
  \psi_m \\
  \psi_{n+1} \\
\end{array}\right),
\end{equation}

\begin{equation}\label{AMTM}
    T_m=\left(%
\begin{array}{cc}
  E-U_m & -1 \\
  1 & 0 \\
\end{array}\right).
\end{equation}
For the second model the state vector can be defined as
\begin{equation}\label{SWv}
v_m=\left(\begin{array}{c}
  \psi_m \\
  \psi^{\prime}_m \\
\end{array}\right),
\end{equation}
where $\psi_m$ and $\psi^{\prime}_{m}$ are the values of the wave
function, $\psi(x)$, and its derivative at the $m$-th interface
between the layers. The transfer matrix in this case takes the
form
\begin{equation}\label{SWTM}
    T_m=\left(%
\begin{array}{cc}
  \cos (k_md_m) & (1/k_m)\sin (k_md_m) \\
  -k_m\sin (k_md_m) & \cos (k_md_m) \\
\end{array}%
\right),
\end{equation}
where $k_m=k\sqrt{\epsilon_m}$. The most important property of the
transfer matrices is that the transfer matrix ${\bf T_M}$
describing the evolution of the initial state vector across the
$M$ sites (slabs) is equal to the product of the one-step matrices
\begin{equation}
{\bf T_M}=\prod _1^M T_m. \end{equation} Using the transfer
matrices, we calculate the finite size LE, which  for both models
is defined as
\begin{equation}\label{LEdef}
\tilde{\gamma}=\frac{1}{L}\ln\frac{\norm{{\bf
T_M}v_0}}{\norm{v_0}},
\end{equation}
where $L$ characterizes the total length of the system. For the
Anderson model, $L=M$ if the distance between adjacent sites is
chosen as a unit of length, and for the wave equation, $L$ is a
sum of the lengths of all slabs, and is  a random quantity.

We calculate LE iteratively using Eq.~(\ref{LEdef}) starting with
an arbitrary initial vector $v_0$. The resultant vector is
re-normalized after every ten iterations in order to avoid any
loss of accuracy.\cite{random matrix} Since we are interested in
statistics of finite size LE, we do not try to find its limiting
value for $L\rightarrow\infty$. Instead, we keep the size of the
system fixed while calculating $\tilde{\gamma}$ for different
realizations of our systems. At the same time, since we are
interested in asymptotic properties of the distribution, we
consider only sufficiently long systems, for which $L\gg l_{loc}$,
where the localization length, $l_{loc}$, is defined through the
average value of LE as
$l_{loc}=\langle\tilde{\gamma}\rangle^{-1}$.

Another quantity of interest in this work is the length $l_s$,
which is expressed in terms of the integral density of states
$N(E)$, Eq.~(\ref{ls}). For the Anderson model $N(E)$ can be
computed with the help of the node-counting
theorem.\cite{node-counting} Starting with an arbitrary initial
vector and the energy values, $E<-2-U$, which are certainly
outside of the energy spectrum of the system, we counted how many
times the sign of the wave function changes over the length of the
system for different values of $E$.  Each new node corresponds  to
a new  state of the system.\cite{node-counting}

For the random superlattice model we find it more convenient to
use the phase formalism described, for instance, in
Ref.~\onlinecite{LGP}. Within this formalism the density of state
is expressed in terms of the phase variable,
$\phi=\tan^{-1}{\left(\psi^{\prime}/\psi\right)}$. In the case of
systems with a single band spectrum, this phase changes between
$0$ and $\pi$ when $E$ sweeps the spectrum of the system from one
band boundary to the other. In the superlattice, the spectrum of
the wave  in the absence of disorder consists of multiple bands.
In this case,  the phase increases by $\pi$ across every allowed
band, and stays constant and equal to $n\pi$, inside any $n$-th
forbidden band. If disorder in our model is not too strong, the
regions of the constant phase are preserved even in the presence
of random fluctuations Fig.~(\ref{fig:phase}), and can be used for
identifying the fluctuation boundaries of the bands in the
disordered system. Then we can introduce a  density of states
$N(E)$ for a \emph{single band}, which is normalized to change
from $0$ to $1$, when energy spans from one fluctuation boundary
to another. $N(E)$ normalized this way is substituted in
Eq.~(\ref{ls}) in order to calculate $l_s$ for the superlattice
model. When disorder becomes stronger the regions of constant
phase disappear, and the notion of the single band density of
states becomes meaningless. In our calculations we always make
sure to avoid such situations.
\begin{figure}
\includegraphics[width=3in,angle=-90]{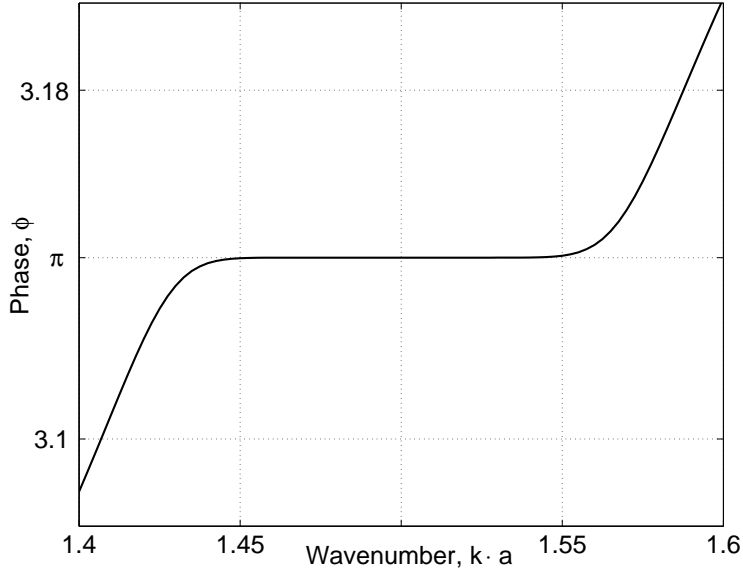}
\caption{The dependence of the phase near the band gap region
($1.44<ka<1.56$) separating the first and the second bands in the
superlattice model. $d_2$ was taken from a uniform distribution
with $\delta$=0.1, $L/a\simeq 10^6$. } \label{fig:phase}
\end{figure}

\section{Scaling description of the moments of the distribution function}

It was shown in Refs.~\onlinecite{DeychPRL2000,DeychPRB} that the
variance, $\sigma^2$ of the Lyapunov exponent in the LLoyd model
can be conveniently described in terms of a relationship between
two scaling variables, $\tau$ defined as
\begin{equation}\label{tau_def}
\tau=\frac{\sigma^2L}{\gamma}
\end{equation}
and $\kappa$, defined as
\begin{equation}\label{kappa}
\kappa=\frac{l_{loc}}{l_s}
\end{equation}
In this paper we show that the variance of LE in more generic
models can also be described in terms of the scaling function
$\tau(\kappa)$.
\begin{figure}
\includegraphics[width=3in,angle=-90]{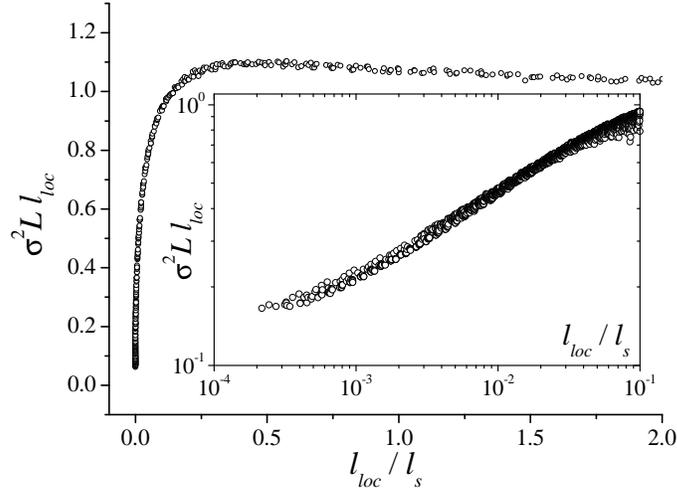}
\caption{Typical dependence of the scaling parameter $\tau$ on
$\kappa$ for the Anderson model. The width of the distribution of
disorder changes from $U=0.08$ to $U = 0.16$. Curves corresponding
to different values of the width are not distinguishable. In the
inset the region of small $\kappa$ is shown in the log-log scale.}
\label{fig:scalingAM}
\end{figure}
\begin{figure}
\includegraphics[width=3in,angle=-90]{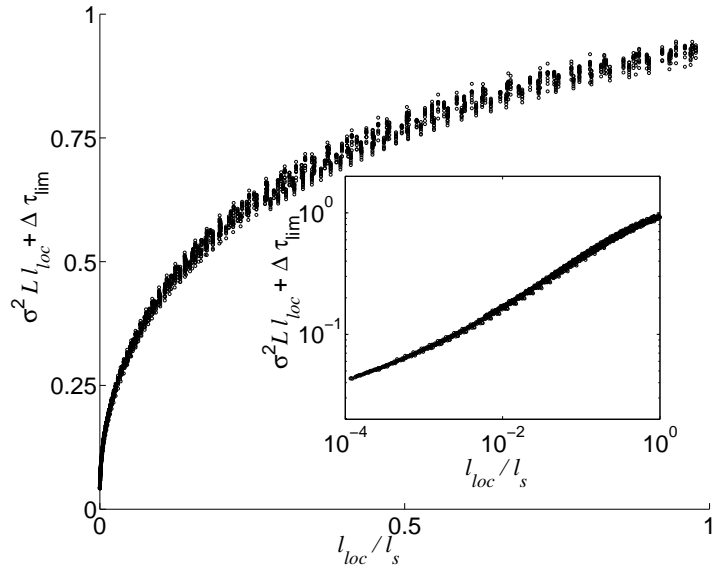}
\caption{Dependence of the scaling parameter $\tau$ on $\kappa$
for the dichotomic distribution (superlattice model) of $d_2$ with
$\delta$ = 0.1, 0.125, 0.15, 0.175, and 0.2. For every value of
disorder we took 17 lengths $L$, ranging from 320 to 20000 layers.
Different offset value of $\tau_{lim}$ was compensated (see text).
In the inset the same is shown  in the log-log scale.}
\label{fig:scalingSL}
\end{figure}

In order to demonstrate this result we computed $\sigma^2$ and
$l_s$ for different values of the energy, strength of disorder,
and length of the system for both models under considerations. The
results of these calculations were presented in the form of the
function $\tau(\kappa)$, which is shown in
Figs.~\ref{fig:scalingAM} and \ref{fig:scalingSL} for the Anderson
model and the superlattice model, respectively. The data included
in these figures correspond to systems with $L\gg l_s,l_{loc}$.
The first important result revealed by this figures is that all
the data lie on a single curve, when expressed in terms of the
variables $\tau$ and $\kappa$ for both models. This result
confirms our general conjecture that the second moment of the
distribution function of LE can be universally described in terms
of variables $\tau$ and $\kappa$ regardless the microscopical
nature of the models under consideration. While the form of the
function $\tau\left(\kappa\right)$, may differ for different
models, its essential qualitative properties show a degree of
universality: $\tau\left(\kappa\right)=1$ for $\kappa>1$, and it
steeply decreases for $\kappa\ll 1$. We are most interested here
in the latter region, where the fluctuation states arise. For the
Lloyd model $\tau=\left(\pi/2\right)\kappa$ for $\kappa\ll
1$,\cite{DeychPRL2000,DeychPRB} while in the models studied in
this paper the dependence of $\tau$ upon $\kappa$ is much steeper.
In order to obtain a better insight into the properties of
$\tau\left(\kappa\right)$ for small $\kappa$, we re-plotted our
numerical data  in log-log coordinates (see insets in
Figs.~\ref{fig:scalingAM} and \ref{fig:scalingSL}). Before
interpreting these figures we have to note that unlike the case of
the Lloyd model, where $\tau(0)=0$, in the models considered here
$\tau(0)$, while very small, is not equal to zero. The reason for
this is the small fluctuations of the LE due to non-resonance
tunnelling through a random barrier, which contributes to $\tau$
at the fluctuation spectrum boundary where $\kappa=0$. This small
contribution is model specific, and in the Anderson model it can
be neglected everywhere with exception of a small neighborhood of
the fluctuation spectrum boundaries. This can be seen from the
fact that while $\kappa$ changes over at least two orders of
magnitude, the data for the Anderson model (inset in
Fig.~\ref{fig:scalingAM}) form a straight line with exceptions of
a few points corresponding to very small values of $\kappa$.
According to these results, $\tau(\kappa)$ has the form
\begin{equation}\label{tau(kappa)}
\tau=C\kappa^{\alpha}+\tau_{lim},
\end{equation}
where $\tau_{lim}$ stands for the non-universal correction
discussed above. In the superlattice model the value of
$\tau_{lim}$ is more significant, and therefore has to be
compensated. In order to estimate coefficients $C$ and $\kappa$,
we select only those data for which $l_s<L$ and use linear
regression. The results of the fit are presented in the table
below:

\vskip 3pt

\begin{center}
\begin{tabular}{|c|c|c|}
  \hline
   & Anderson model & Superlattice models \\
  \hline
  $ C $ & $1.27$ & $1.08$ \\
  $ \alpha $ & $0.27$ & $0.40$ \\
  \hline
\end{tabular}
\end{center}
These results demonstrate that while the nature of the scaling
parameters is universal for both models, the numerical values of
the respective parameters are model dependent. An interesting
question is whether the values of $C$ and $\alpha$ depend upon the
type of statistics of the respective random parameters of our
models (site energy for Anderson model, the layer width for the
superlattice model). In the case of a superlattice model we found
that the change in statistics (from the box to dichotomic
distribution) did not affect the values of the coefficients $C$
and $\alpha$. For the Anderson model with the dichotomic
distribution of the site energies the results were inconclusive.
Strong noise in the data for the dichotomic process prevented us
from positively establishing equivalency of the coefficients for
the two different types of statistics.

In the region of fluctuation states, a new intermediate regime of
lengths $L$, in which $l_{loc}\ll L\ll l_s$ appears. This regime
does not exist for in-band states. It is natural to anticipate
that the scaling behavior of our systems in this regime would
change. In order to study this question, we divided our data in
groups according to the value of $L/l_s$, including points with
$L/l_s>1$ as well as with $L/l_s<1$. Carrying out statistical
analysis of the data for fixed values of $L/l_s$ we were able to
obtain dependencies of the parameters $C$ and $\alpha$ on $L/l_s$;
the respective results are presented in Figs.~\ref{fig:CalphaAM}
and \ref{fig:CalphaSL}. First of all, we would like to note that
these dependencies saturate to values presented in the Table $1$
for $L/l_s>1$. This confirms our assumption that in this regime
$\tau$ depends upon a single parameter, $\kappa$.
\begin{figure}
\includegraphics[width=3in,angle=-90]{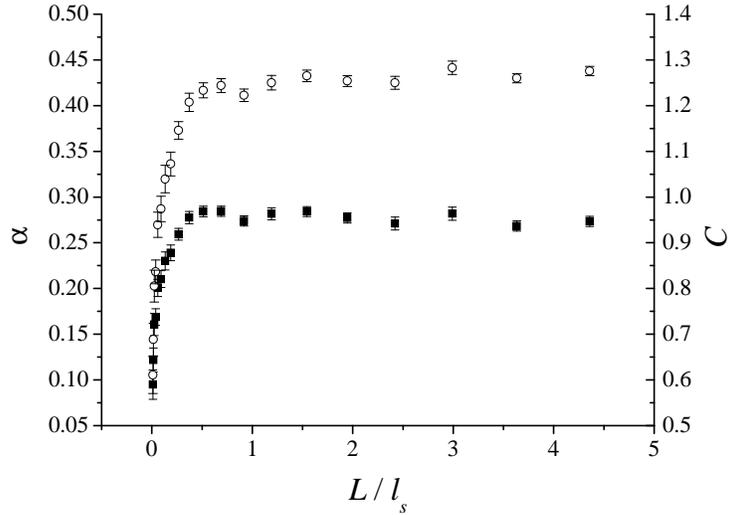}
\caption{Dependence of the index of the scaling parameter $\alpha$
(filled squares, left axis) and the factor $C$ (circles, right
axis) on $L/l_s$ for the Anderson model.} \label{fig:CalphaAM}
\end{figure}
\begin{figure}
\includegraphics[width=3in,angle=-90]{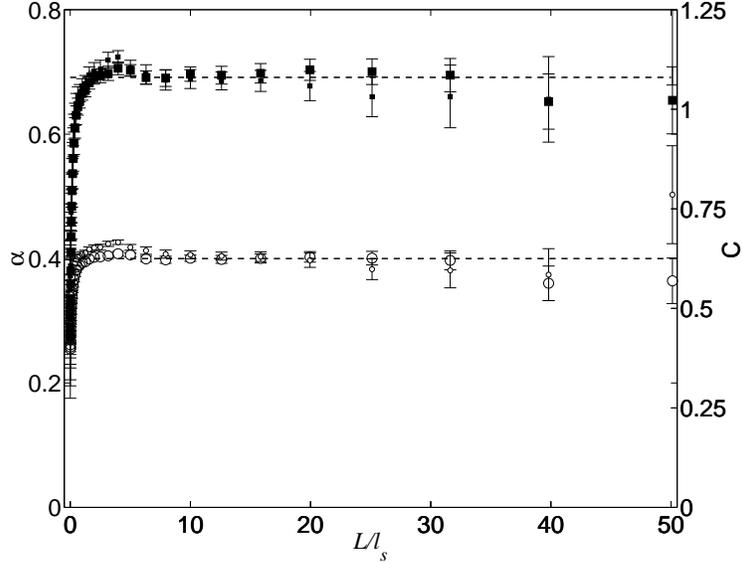}
\caption{The exponent of the scaling parameter $\alpha$ (filled
squares, left axis) and the factor $C$ (circles, right axis) as
functions of $L/l_s$ for the superlattice model. Large and small
symbols correspond to dichotomic and box distribution of $d_2$,
respectively.} \label{fig:CalphaSL}
\end{figure}

For shorter systems, however, a new parameter, $L/l_s$ emerges.
For the Anderson model we were able to show that $\alpha(L/l_s)$
is best described by the logarithm $\alpha(L/l_s)\sim
\ln\left(l_s/L\right)$, which means that the variance of the
Lyapunov exponent, $\sigma^2$, in this regime demonstrates an
anomalous scaling with the length of the system, $L$:
\begin{equation}\label{sigmaanom}
\sigma^2 \propto \frac{1}{Ll_{loc}}\exp{\left[\alpha(L/l_s)\ln{\kappa}\right]%
}\propto L^{-(1+\ln\kappa)}.
\end{equation}
It is interesting to note that when $\kappa$ decreases,
$1+\ln\kappa$ may become negative resulting in $\sigma^2$
increasing with $L$. This behavior can be qualitatively understood
from the following arguments: The condition $L\ll l_s$ means that
for the most of the realizations of the random potential no states
exist in the energy interval under discussion. The transmission
through such realizations fluctuates rather weakly. The greatest
contribution to the transmission fluctuations is given by those
few realizations that can support at least a single state. The
probability for such realizations to arise grows when the length
of the system increases, resulting in the respective increase of
$\sigma^2$. This behavior, of course, breaks down for very large
values of $l_s$, which correspond to states close to the genuine
spectral boundary, because for these states $\sigma^2$ is
determined by a non-universal correction to $\tau$ given by
$\tau_{lim}$.

The behavior of $\sigma^2$ given by Eq.~(\ref{sigmaanom}) can be
confirmed by plotting directly the function $\sigma^2(L)$ for
energies from the band-gap. Fig.~\ref{fig:sigmaofAM} presents such
a plot for the Anderson model for the value of $\kappa$ equal to
$\kappa=0.2$. It demonstrates a good agreement with
Eq.~(\ref{sigmaanom}): the slope of the curve was found to be
equal to $1.77$, while an estimate for this slope from
Eq.~(\ref{sigmaanom}) gives $1.78$. It should be noted, however,
that the regime described by Eq.~(\ref{sigmaanom}) exists in a
relatively narrow interval of energies, at least for the Anderson
model with the box distribution. The reason for this is that $l_s$
grows very fast in the region of fluctuation states when the
energy is shifted toward the fluctuation spectrum boundary. Very
large $l_s$ means that only few realizations of our system support
at least a single state. Therefore, for the most realizations
transmission occurs via non-resonant under-barrier tunnelling. The
statistics of the transmission for this subset of realizations is
determined by the localization length alone ($l_s$ is exact
infinity for these realizations). As a result, we have a
competition between a small number of realizations, supporting
states, for which fluctuations of the Lyapunov exponent are large
and grow with the length, and the majority of realizations, in
which $\sigma^2$ is small, and decreasing with length. At very
large $l_s$ the contribution to $\sigma^2$ from the representative
realizations becomes larger than the contribution from the
resonant realizations, and Eq.~(\ref{sigmaanom}) fails. In this
case, an asymptotic behavior of $\sigma^2$ is again controlled by
the localization length alone, as it can be seen in
Fig.~\ref{fig:scaling-off}, where $\sigma^2L$ saturates at $L$
much smaller than $l_s$.
\begin{figure}
\includegraphics[width=3in,angle=-90]{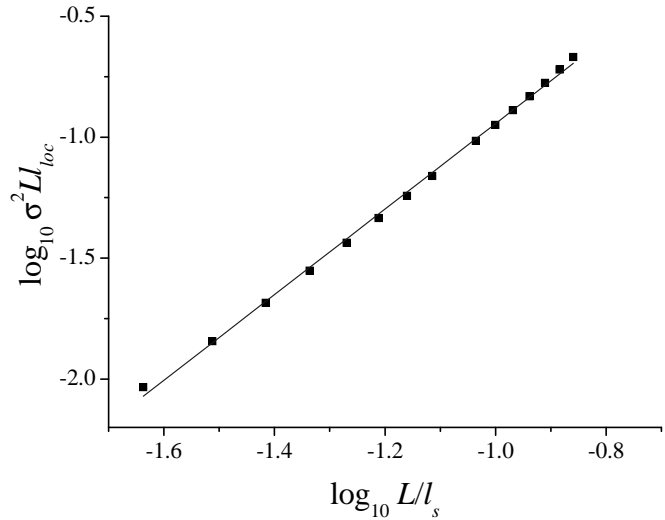}
\caption{The logarithm of the scaling parameter $\tau$ for the
Anderson model as a function of $\log_{10}L/l_s$ for intermediate
values of energy when $l_s$ is not too large. Points are the
result of numerical calculations and the straight line is a linear
fit.} \label{fig:sigmaofAM}
\end{figure}

\begin{figure}
\includegraphics[width=3in,angle=-90]{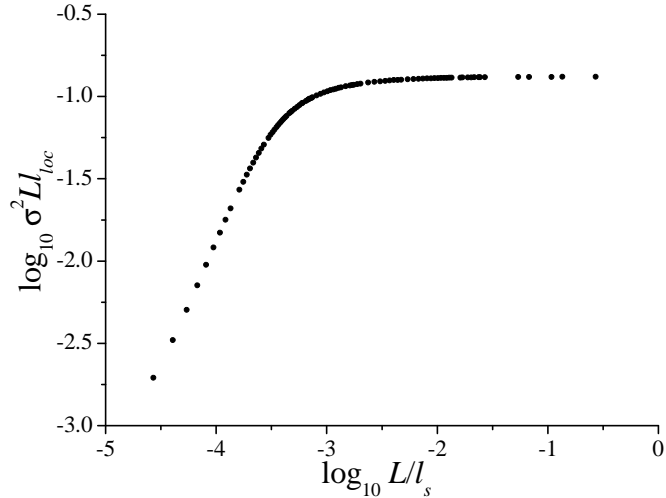}
\caption{The logarithm of the scaling parameter $\tau$ for the
Anderson model as a function of $\log_{10}L/l_s$ for energies
corresponding to extremely large values of $l_s$. The saturation
occurs at the length close to the localization length.}
\label{fig:scaling-off}
\end{figure}

The assumption about the Gaussian form of the distribution of LE
is the result of the central limit theorem, and strictly speaking
is true only asymptotically when $L\rightarrow\infty$. At finite
$L$ the distribution function deviates from the Gaussian form even
in the regime when SPS holds.\cite{Abrikosov,DPMK,Mello} However,
it was found in Refs.~\onlinecite{Schomerus,DeychPRL2003} that
this deviation, as measured by the magnitude of the third and
higher moments, increases significantly in the vicinity of the
band boundary of the initial spectrum. This result was obtained
analytically for the white-noise potential in
Ref.~\onlinecite{Schomerus}. The first study of the scaling
properties of the third moment was reported in
Ref.~\onlinecite{DeychPRL2003}. In this part of the paper we
expand scaling analysis of Ref.~\onlinecite{DeychPRL2003} to the
superlattice model, and compare the results obtained for these two
models. We consider the scaling properties of the third cumulant
$\varrho = \langle (\gamma -\langle\gamma\rangle)^3\rangle$, which
characterizes the asymmetry or skewness of the distribution
function. Fig.~\ref{fig:skewnessAMin} shows the energy dependence
of the third moment for the Anderson model. It is seen that this
moment significantly grows in the vicinity of the initial band
boundaries of both models, which means that the significant
deviation of the distribution function of LE from the Gaussian
form in the region, where traditional SPS violates is a universal
phenomenon.
\begin{figure}
\includegraphics[width=3in,angle=-90]{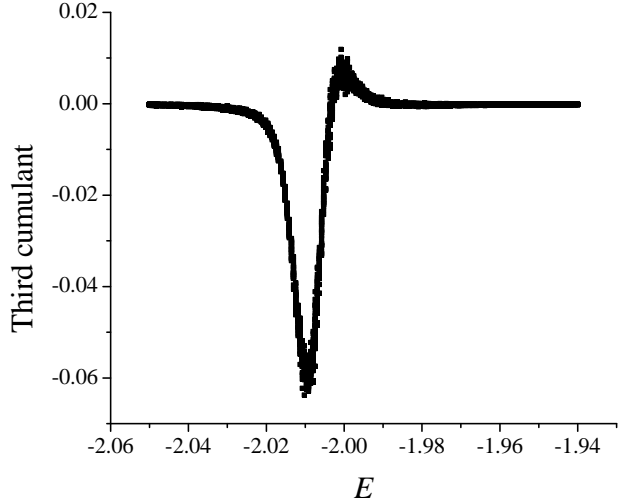}
\caption{Dependence of the renormalized third cumulant, $\varrho
L^2$, on energy in the vicinity of the band edge of a pure system
($U=0.05$) for the Anderson model.} \label{fig:skewnessAMin}
\end{figure}

To analyze scaling properties of the third cumulant we consider
the dimensionless parameter
\begin{equation}  \label{eq:skewness}
\tau_3 = \varrho L^2 l_{loc}
\end{equation}
The dependence of $\tau_3$ on $\kappa$ for the Anderson model and
superlattice is shown in Figs.~\ref{fig:skewnessAM} and
\ref{fig:third_cumulant_scaling}, respectively.
\begin{figure}
\includegraphics[width=3in,angle=-90]{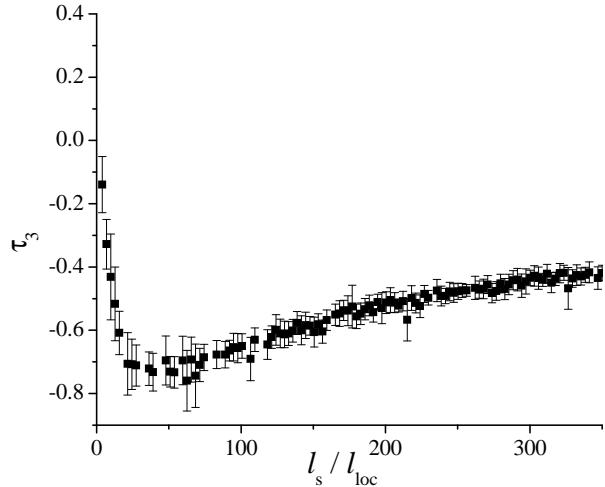}
\caption{Dependence of the parameter $\tau_3 = \varrho L^2
l_{loc}$ on $\kappa^{-1} = l_s/l_{loc}$ (Anderson model) for a set
of different widths of the distribution of the potential: $0.001 <
U < 0.21$. Error bars show the dispersion of the results of
numerical simulations near a mean values shown by squares.}
\label{fig:skewnessAM}
\end{figure}
\begin{figure}
\includegraphics[width=3in,angle=-90]{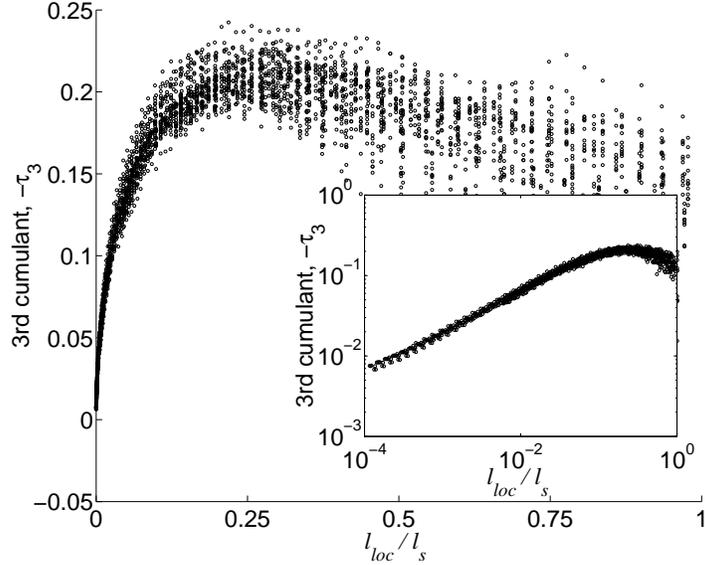}
\caption{Dependence of $\tau_3$ on $l_{loc}/l_s$ for dichotomic
distribution (superlattice model) of $d_2$ with $\delta$ = 0.1,
0.125, 0.15, 0.175, and 0.2. For every value of disorder we took
17 length $L$, ranging from 320 to 20000 layers. On the insert the
same is shown in log-log scale.}
\label{fig:third_cumulant_scaling}
\end{figure}

One can see from these figures that while data for the parameter
$\tau_3$ are rather noisy, it shows a relatively good scaling
behavior as a function of the single parameter $\kappa$ for both
models. This fact itself is quite remarkable since it demonstrates
that even in the region, where the distribution function of LE
deviates significantly from the Gaussian form, it can still be
characterized by two parameters within the scaling procedure
suggested here.

The better data quality for the superlattice model allowed for
more thorough study of the third moment. The insert in
Fig.~\ref{fig:third_cumulant_scaling} shows a good scaling
behavior similar to Eq.~(\ref{tau(kappa)}):
\begin{equation}\label{tau2(kappa)}
-\tau_3=C_3\kappa^{\alpha_3}+\tau_{3,lim}
\end{equation}
The limiting value $\tau_{3,lim}$, was substantially smaller than
$\tau_{lim}$, so no explicit correction was needed to obtain
Fig.~\ref{fig:third_cumulant_scaling}.

For intermediate lengths, $l_{loc}\ll L\ll l_s$, we analyzed data
using approach  similar to that employed to obtain
Figs.~\ref{fig:CalphaAM} and \ref{fig:CalphaSL}. For fixed values
of $L/l_s$ we obtained dependencies of the parameters $C_3$ and
$\alpha_2$ on $L/l_s$ (Fig.~\ref{fig:C2alpha2SL}), and found the
saturated values of $C_3=0.73$ and $\alpha_3=0.52$ -- the same for
both dichotomic and box distributions.
\begin{figure}
\includegraphics[width=3in,angle=-90]{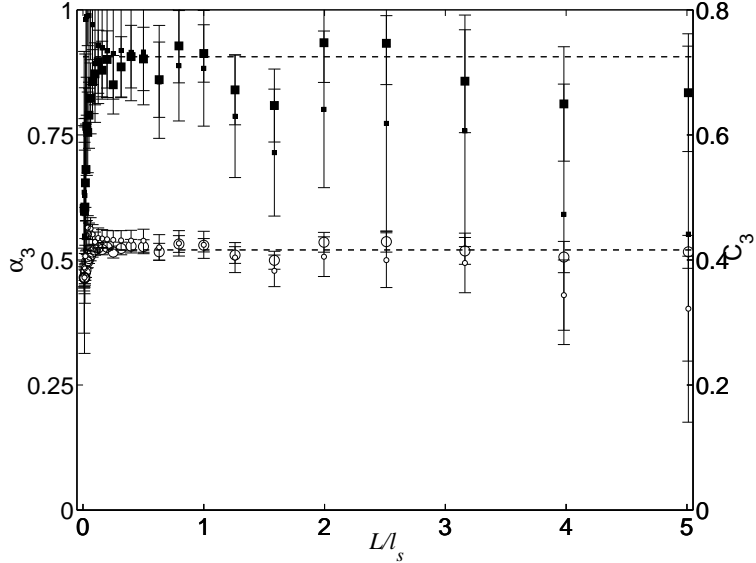}
\caption{$\alpha_3$ (filled squares, left axis), the exponent of
$\tau_3$,  and factor $C_2$ (circles, right axis) as functions of
$L/l_s$ for superlattice model. Large and small symbols correspond
to dichotomic and box distribution of $d_2$ respectively.}
\label{fig:C2alpha2SL}
\end{figure}

\section{Comparison with the Gaussian white noise model}

It is well known that under certain circumstances statistical
properties of one-dimensional disordered systems in the vicinity
of the band edges of the initial spectrum can be universally
described by replacing an actual random potential by a Gaussian
white noise potential.\cite{LGP} One of the manifestations of this
fact is that the statistical properties of LE in the Anderson
model with the box distribution of the site
energies\cite{Derrida,Goldhirsch,Schomerus}  are very similar to
those of the continuous model with the Gaussian white-noise
potential,\cite{LGP} and are characterized by the same scaling
parameter $E/D^{2/3}$, where $E$ is the energy counted from the
initial band boundary, and $D$ is the variance of the random
potential. It was noted in Ref.~\onlinecite{DeychPRB} that  the
scaling parameter $\kappa$ is a single-valued function of the
Gaussian scaling parameter $E/D^{2/3}$ for the white-noise model,
so that in this case these two parameters are equivalent to each
other. An important question now arises:  whether the apparent
universality of the scaling description, suggested in this paper,
is a mere consequence of the fact that in the region of the
fluctuation states all models can be reduced to the Gaussian
model, or this universality reveals more fundamental properties of
this spectral region.  This question was partially discussed in
the Ref.~\onlinecite{COBanomaly}, in which it was shown that the
behavior of the second moment of the LE in the vicinity of $E=0$
of the Anderson model obeys the scaling description in terms of
the parameter $\kappa$, while the Gaussian approximation certainly
does not work in this part of the spectrum. In this paper, we
address this question considering regions of the fluctuation
states in the superlattice model. \footnote{For the Anderson model
our calculations confirmed the fact that the scaling behavior of
this model is equivalent to that of the models with white-noise
potential.} The inset to Fig.~\ref{fig:Gaussscaling} shows the
plot of the  parameter $\tau$ versus the Gaussian scaling
parameter $(k-k_i)/D^{2/3}$ ($D\propto \delta^2$), where $k_i$ is
the dimensionless frequency of one of the initial band boundaries
of the superlattice for several values of the disorder. Moreover,
we included the frequencies from the upper edge of the first band
and the lower edge of the second band. We found that instead of
$E/D^{2/3}$ predicted by the Gaussian white noise model, our data
are better scaled with the parameter $E/D^{1/2}$. One can see from
Fig.~\ref{fig:Gaussscaling}, that while the Gaussian scaling
fails, the function $\tau(\kappa)$ discussed in the previous
section of the paper gives the best scaling description of this
model as well as of the Anderson model. We can conclude,
therefore, that the scaling parameter $\kappa$ retains its
universal significance beyond the validity of the white-noise
approximation.
\begin{figure}
\includegraphics[width=3in,angle=-90]{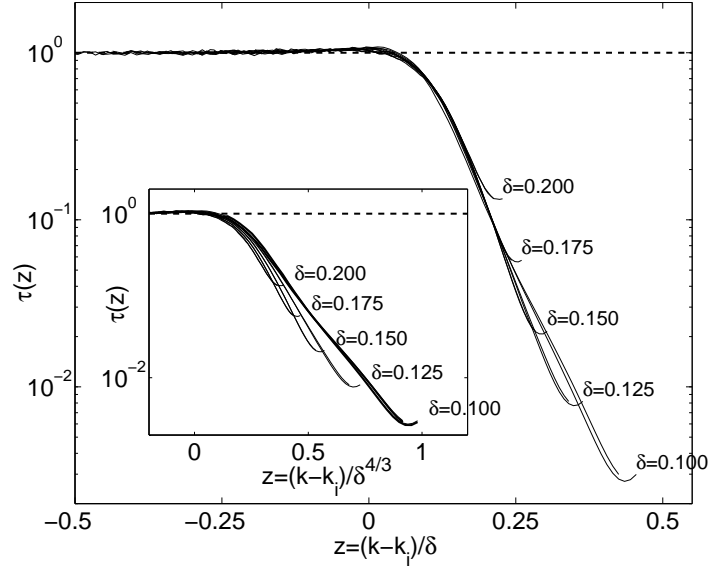}
\caption{Normalized variance of LE, $\tau$, plotted versus
parameter $(k-k_i)/\delta$ demonstrates a good scaling. Scaling
with $(k-k_i)/\delta^{4/3}$, predicted by the Gaussian white noise
model, shown in the inset, fails. The data was generated in the
superlattice model with the box distribution for five values of
$\delta=$0.1, 0.125, 0.15, 0.175, and 0.2, $L$=20000. We included
the frequencies from two band edges - the upper edge of the first
band and the lower edge of the second band. Altogether, the band
gap region between the first and the second bands is covered
entirely.} \label{fig:Gaussscaling}
\end{figure}

\section{Conclusion}

In this paper we studied scaling properties of the distribution
function of the Lyapunov exponent for two one-dimensional
disordered models: the Anderson model with diagonal disorder, and
the model of a scalar wave propagating in a random superlattice.
The main result of the paper is that in the region of band-edge
and fluctuation states, where simple SPS fails, the distribution
function can be described by two independent scaling parameters:
the localization length, $l_{loc}$, and an additional length
$l_s$, introduced in Refs.~\onlinecite{DeychPRL2000,DeychPRB},
which is related to the integral density of states. The fact of
principal importance is that not only the second moment of the
distribution is described by these two parameters, but so also is
the third moment. This means that even though in the region of
fluctuation states the form of the distribution function strongly
deviates from the Gaussian, it still can be described within the
suggested two-parameter scaling approach.

Among other important results of the paper we would like to note
the detailed study of the properties of the variance and the third
moment of LE in the region of fluctuation states. We showed that
both, the normalized variance and the third cumulant presented by
the scaling functions $\tau$ and $\tau_3$, demonstrate a power law
dependence upon the scaling parameter $\kappa$. Parameters of this
power law dependence were found to depend weakly upon the type of
statistics used to characterize our random systems, but are
different for the Anderson model and the superlattice model. When
the length of the system becomes smaller than $l_s$, we showed
that the scaling behavior of $\sigma^2$ deviates significantly
from the central limit theorem behavior even when $L$ remains much
bigger than the localization length.

\end{document}